\documentclass[11pt]{article}
\usepackage{blois2002,epsfig}
\def\be{\begin{equation}}
\def\ee{\end{equation}}
\def\bea{\begin{eqnarray}}
\def\eea{\end{eqnarray}}

\newcommand\noi{\noindent}
\newcommand\etal{{\it et al}.\ }
\newcommand\etalcom{{\it et al}.,}

\newcommand\ep{\epsilon}
\newcommand\sintb{\sin2\beta}

\newcommand\notcp{\not{\!\!\!\!CP}}

\newcommand{\lsim}{\mathrel{\lower4pt\hbox{$\sim$}}
\hskip-12.5pt\raise1.6pt\hbox{$<$}\;}

\newcommand{\gsim}{\mathrel{\lower4pt\hbox{$\sim$}}
\hskip-12.5pt\raise1.6pt\hbox{$>$}\;}
\begin{document}

\begin{flushright}
BNL-HET-02/29 \\
AMES-HET 02-08
\end{flushright}

\vspace*{1cm}

\title{ PRISTINE DETERMINATION OF THE UNITARITY TRIANGLE USING\\ 
B $\to$ K $D^0$ PROCESSES}

\author{David Atwood$^{a\dag}$ and Amarjit Soni$^{b\ddag}$
\bigskip
}
\author{
$^\dag$
Department of Physics and Astronomy\\
Iowa State University\\ 
Ames, IA\ \ 50011
\medskip
}
\author{
$^\ddag$
High Energy Theory Group, Physics Department\\
Brookhaven National Laboratory\\ 
Upton, NY\ \ 11973
\medskip
}
\author{$^a$atwood@iastate.edu
\medskip
}
\author{$^b$presenter;  soni@bnl.gov
\bigskip
}

\maketitle

\abstracts{
There is no good reason to think that BSM-CP-odd phases(s) will
necessarily cause large deviations in B-physics from predictions of the
SM. Therefore, residual theory error in extraction of the unitarity
triangle can undermine experimental efforts to search for BSM phase(s). We
stress that final states containing $D^0$ or $\bar D^0$ in decays of
charged {\it and} neutral B's can yield all the angles of the unitarity
triangle with negligble theory
error (i.e. $O(0.1\%)$).
}

\section{Introduction \& Motivation: 
why must we target unitarity triangle with ``zero" 
theory error}

The two $B$-factories have made
considerable progress in determining $\beta$ by measurement of the time
dependent CP asymmetry in $B^0(\bar B^0)\to\psi K_s$ and related
modes\cite{higu,raven}. These machines have also been performing remarkably well. Very
soon each of them should have $\sim10^8$ $B$-$\bar B$ pairs and with
improvements in luminosity to $\gsim 10^{34}$ cm$^{-2}$s$^{-1}$  that
are anticipated, $10^9$ $B$-$\bar B$ 
pairs should become accessible in the next few years. In addition,
Tevatron experiments CDF/D$\emptyset$, and hopefully in the not too
distant future $B$TeV and LHCB should also enable much larger data samples.
Furthermore, encouraged by the success of the two B-factories, there is also 
considerable interest in very high luminosities ($\approx 10^{36} cm^{-2} sec^{-1}$)
facilities, Super-BELLE and Super-BABAR.
With these anticipated experimental developments it is not enough that
we can determine $\beta$ with essentially negligible theory error
(actually $\lsim1\%$) we must target $\alpha$ and $\gamma$ extraction
also with zero theory error\cite{zero}. Indeed extremely accurate determinations
of all 3 unitarity angles is not just desirable but may well be an
essential prerequisite for a successful search of the effect of any
CP-odd phase in B-decays due to physics beyond the Standard Model.  

In this regard it is important to realize that although the CP-odd
phase in the CKM picture is of $O(1)$ and not small and it leads to
large asymmetries in $B$-physics, it yields very small CP-asymmetries
in $K$ decays; recall $\ep_K\sim10^{-3}$ and $\ep^\prime\sim10^{-6}$ even
though $a_{CP}(B\to\psi K_s)\sim 75\%$. 
There are two important repercussions of this realization.

\begin{enumerate}

\item Failure of the ($b \to d$) unitarity triangle [UT] due to effects of new
physics may well be small and subtle. Therefore, residual theory error
in the determination of the angles of the UT may mask the effect of 
new physics 
and thwart experimental attempts to find them.

\item Search for effect of any BSM phase(s) may well require very large
data samples. A model independent estimate is very difficult to make.
If the asymmetry due to new physics is of order $\ep_K\sim10^{-3}$,
then even with a branching ratio of $\sim10^{-3}$ (there are very few
relevant $B$-decay modes that have branching ratios this big)  we may
need $\sim10^{10}$ $B$'s to find such an effect.  Efforts at developing
the capabilities for large data samples of $B$'s in clean environments and/or
specialized $B$-detectors are therefore very worthwhile.

\end{enumerate}

\noi These considerations lead us to suggest that while attempts at
using proposed methods for $\alpha $ and $\gamma$ via $\pi\pi$,
$\rho\pi$, $K\pi$ etc.\ \cite{kpi} should continue, we should realize that the
presence of penguin, especially 
electroweak penguins (EWP), contributions along with the use
of flavor symmetries could easily lead to theory errors of 10\% or more
due to any model dependence and theoretical assumptions that need be
invoked. Clearly on both the theory as well as the experimental front
methods for determining\cite{notation} $\alpha$, $\beta$, $\gamma$
with zero theory
error should be vigorously pursued.

Recall that the current measurements
of $\sintb$ agree very well with theoretical
expectations; the experimental world
average $\sintb^{WA}=0.78\pm.08$\cite{higu,raven,affold} is completely consistent
with theoretical 
expectations,  $(\sintb)^{SM}\approx 0.70\pm.10$\cite{ciuch,hocker,atwoodone}.
See Table~\ref{tabone} for a comparison of some theoretical fits.
It is important though to realize that this test
of the SM has serious limitations. A lot of theoretical
input is used in making these fits as they rely heavily on theoretical
evaluations of several of the hadronic matrix elements. Improving
the accuracy in these calculations is extremely difficult
and painfully slow. To underscore this we mention 
two problems with the theoretical input being used
in these fits.

\begin{table}[t]
\caption{Comparison of some fits.\label{tabone}}
\hspace*{-.5in}\begin{tabular}{|c|c|c|c|}
\hline
Input Quantity 
& Atwood \& Soni\cite{atwoodone}
& Ciuchini {\it et al}\cite{ciuch}
& Hocker {\it et al}\cite{hocker} 
\\
\hline
\hline
$R_{uc} \equiv |V_{ub}/V_{cb}|$ & $.085\pm.017$ & $.089\pm.009$ &
$.087\pm.006\pm.014$ \\
$F_{B_d} \sqrt{\hat B_{B_d}}$ MeV & $230\pm50$ & $230\pm25\pm20$ &
$230\pm28\pm28$ \\
$\xi$ & $1.16\pm.08$ & $1.14\pm.04\pm.05$ & $1.16\pm.03\pm.05$ \\
$\hat B_K$ & $.86\pm0.15$ & $.87\pm0.06\pm0.13$ & $.87\pm.06\pm.13$
\\
\hline
Output Quantity & & & \\
\hline
$\sin2\beta$ & $.70\pm.10$ & $.695\pm.065$ & $.68\pm.18$ \\
$\sin2\alpha$ & $-.50\pm.32$ & $-.425\pm.220$ & \\
$\gamma$ & $46.2^\circ\pm9.1^\circ$ & $54.85\pm6.0$ & $56\pm19$ \\
$\bar \eta$ & $.30\pm.05$ & $.316\pm.040$ & $.34\pm.12$ \\
$\bar \rho$ & $.25\pm.07$ & $.22\pm.038$ & $.22\pm.14$ \\
$|V_{td}/V_{ts}|$ & $.185\pm.015$ & & $.19\pm.04$ \\
$\Delta m_{B_s}(ps^{-1})$ & $19.8\pm3.5$ & $17.3^{+1.5}_{-0.7}$ &
$24.6\pm9.1$ \\
$J_{CP}$ & $(2.55\pm.35)\times10^{-5}$ & & $(2.8\pm.8)\times10^{-5}$ \\
$BR(K^+\to\pi^+\nu\bar\nu)$ & $(0.67\pm0.10)\times10^{-10}$ & &
$(.74\pm.23)\times10^{-10}$ \\
$BR(K_L\to\pi^0\nu\bar\nu)$ & $(0.225\pm0.065)\times10^{-10}$ & &
$(.27\pm.14)\times 10^{-10}$ \\
\hline
\end{tabular}
\end{table}

First, even for the highly matured calculation of
$B_K$ there are some reasons to believe that the JLQCD (quenched)
result,  $\hat B^{Staggered}_K=0.860\pm.058$\cite{aokione}, 
which has been widely used in the past many years,
may well be 
$10-15 \%$ higher than the true value. This expectation is based
on results obtained by using the newer discretization,
domain wall quarks (DWQ) which has much better chiral-flavor symmetry
properties. With DWQ both CP-PACS\cite{ali} and RBC\cite{blumone} 
get smaller
values; averaging their numbers
one gets $\hat B^{DWF}_K =
0.758\pm0.033$; (again this is quenched).
This is about $\simeq 13\%$ below the older
JLQCD result. 

Second problem is with the SU(3) breaking ratio, $\xi$, which monitors
$B_s$ versus $B_d$ oscillations.
The concern with regard to $\xi$ that we have been voicing in the past
couple of years \cite{atwoodone,sea1,moriond1} is that the widely used central value
($\sim 1.15$) provided by some lattice calculations, is quite
likely an underestimate and even more importantly the quoted error on
$\xi$ of
$\sim0.05$ appears too low.
This worry is based on the observation that most lattice calculations
of $\xi$ use the ``indirect" method in which the matrix
element of the 4-quark operator is parmetrized in terms of a B-parameter
and the pseudoscalar decay constant. In actual numerical
calculations of the decay constant, the usual practice is
to fit linearly 
to the light quark mass ($m_q$) dependence.
Since the orginal matrix element of interest depends
quadratically on the decay constant, this procedure is unlikely
to get the right coefficient, for example, of $m_q^2$.
Infact, these matrix elements can be calculated directly
on the lattice\cite{bern,lell}. There is no need to introduce B-paramerters.
Use of the B-parameters in calculating the $\Delta F=2$
mixing matrix elements is purely a historical accident.
Recall that originally B-parameters were introduced
in such calculations
for dealing with the analogous $K - \bar K$ mixing matrix element.
There $f_K$ was known experimentally and in the phenomenological
literature $B_K$ was introduced as a measure of deviation of
the matrix element for the idealised case of vacuum saturation.
For B-mesons, the decay constant is not known from experiment
and determining that from theory becomes the central issue.
One can equally well directly calculate (on the lattice) 
the mixing matrix element without introducing
decay constant or B-parameters\cite{bern,lell}. The direct method seems
to give largish central value but within (rather large)
errors is consistent with the indirect method.
To be on the safe side one ought to use both methods 
with very good control over errors in each case and then an average
of the two methods should be used for $\xi$. 

Meantime, in light of this observation, we \cite{atwoodone} adopt a
conservative attitude and had used $\xi=1.16\pm0.08$ with an error that is
considerably  bigger compared to Hocker \etal\ \cite{hocker} and Ciuchini \etal
\cite{ciuch}; and we have been stressing for quite sometime
that their errors are an underestimate\cite{sea1,moriond1}. 
We
also examined the effect on $\sintb$ of an increase in $\xi$ (including
a larger error) along with a decrease
in $B_K$, following indications from DWQ. (See Table~\ref{tabtwo}
from \cite{moriond1}).
With the size of uncertainties currently present, $\sintb$ is hardly
effected with $(\sintb)^{SM}=0.72\pm.10$ giving us additional
confidence in the comparison 
of the experimental measurements to the
predictions of the SM\null. However, the point still remains
that use of theory input has its limitations and we must
try hard to develop methods that can yield 
angles of the unitarity triangle very ``cleanly'', i.e.\ with
zero theory error and with no theoretical assumptions.

\begin{table}[t]
\caption{Stability of our Fit.\label{tabtwo}}
\begin{center}
\hspace*{-.5in}\begin{tabular}{|c|c|c|c|}
\hline
Input Quantity & Atwood \& Soni \cite{atwoodone} &  & \\
\hline
$R_{uc} \equiv |V_{ub}/V_{cb}|$ & $.085\pm.017$ &  &
 \\
 $F_{B_d} \sqrt{\hat B_{B_d}}$ MeV & $230\pm50$ MeV &  &
  \\
  $\xi$ & $1.16\pm.08$ &  & $1.25\pm.10$ \\
  $\hat B_K$ & $.86\pm0.15$ & $.75\pm.13$  & $.75\pm.13$
  \\
  \hline
  Output Quantity & & & \\
  \hline
  $\sin2\beta$ & $.70\pm.10$ & $.73\pm.10$ & $.72\pm.10$ \\
  $\sin2\alpha$ & $-.50\pm.32$ &  & \\
  $\gamma$ & $46.2^\circ\pm9.1^\circ$ & $48.7\pm8.5$ & $52.3\pm12.1$ \\
  $\bar \eta$ & $.30\pm.05$ & $.32\pm.05$ & $.33\pm.05$ \\
  $\bar \rho$ & $.25\pm.07$ &  &  \\
  $|V_{td}/V_{ts}|$ & $.185\pm.015$ & &  \\
  $\Delta m_{B_s}(ps^{-1})$ & $19.8\pm3.5$ &  &  \\
  $J_{CP}$ & $(2.55\pm.35)\times10^{-5}$ & &  \\
  $BR(K^+\to\pi^+\nu\bar\nu)$ & $(0.67\pm0.10)\times10^{-10}$ & &
   \\
   $BR(K_L\to\pi^0\nu\bar\nu)$ & $(0.225\pm0.065)\times10^{-10}$ & &
    \\
    \hline
    \end{tabular}
    \end{center}
\end{table}

We emphasize here that final states containing $D^0$, $\bar D^0$ in decays of charged or neutral
B's can be used very effectively to determine all 
three angles with essentially zero theory error, i.e. $\lsim 0.1\%$.

$B^\pm$ and $B^0$ decays to (e.g.) $K^\pm$ $ D^0(\bar D^0)$ and $K^0/\bar
K^0$ $D^0(\bar D^0)$ respectively involve only decays via two tree
graphs ($b\to c$ and $b\to u$),
no penguin strong or electroweak are involved. 
Methods based on these 
using direct \cite{atwoodtwo} and time dependent \cite{atwoodthree} CP
respectively can give all three angles with no theory error or theory
assumptions. In the case of $B^\pm\to K^\pm D^0(\bar D^0)$, the
$D^0(\bar D^0)$ decays must proceed to CP-non-eigenstates. Using {\it only\/}
$D^0,\bar D^0$ decays to CP eigenstates as advocated in \cite{gronone} has the 
difficulty that the suppressed branching ratio is not accessible to
experiment as $D^0,\bar D^0$ flavor is difficult to tag
in these B-decays\cite{atwoodtwo}.
Estimating this branching ratio with the use of theory input or
assumptions defeats our original goal of determination of the UT with
zero theory error.  A second undesirable feature of using only
CP-eigenstates of $D^0$ is that the resultant CP asymmetry is small
$\sim0$ (a few \%) \cite{gronone}; with CPNES method \cite{atwoodtwo}
the asymmetries are large.

\subsection{$\gamma$ with zero theory error}

This is a uniquely clean method with no theoretical assumption and
involving no penguin contribution, QCD or EW\null. Interference between
two tree graphs, $b\to u$ and $b\to c$ is exploited. Consequently, the
limiting theory error is completely negligible. Furthermore, the
interference between the amplitudes contributing to common final states
of $D^0,\bar D^0$ that are not CP-eigenstates is large resulting in
large direct CP asymmetry $\sim$ tens of percents which can be studied, in
principle, at any $B$-facility\cite{atwoodtwo}. 
Furthermore, while only two modes are
essential for the analysis, many modes are available. Thus discrete
ambiguity in determination of $\gamma$ can be removed 
by use of several modes.

As a specific example one may consider $B^-\to K^-D^0(\bar D^0)$ with
$D^0(\bar D^0)\to K^+\pi^-$ so the overall reaction being studied is
just $B^-\to K^-K^+\pi^-$. It is important to understand that the
suppressed branching ratio $B^-\to K^-\bar D^0$ is {\it not\/} needed
and in fact is an output, i.e.\ is determined in the analysis along
with $\gamma$.  The FS (e.g.
 $K^-K^+\pi^-$) results from interference between two amplitudes one of
 which is color allowed ($B^-\to K^-D^0$) but doubly-Cabibbo-suppressed
 ($D^0\to K^+\pi^-$) whereas the other is color-suppressed ($B^-\to
 K^-\bar D^0$) but Cabibbo allowed ($\bar D^0\to K^+\pi^-$).
 
$B^-$ and $B^+$ decay amplitudes to two such final states (say, e.g.
$B^\pm\to K^\pm D^0,\bar D^0$ with $D^0,\bar D^0\to K^\pm \pi^\mp$ and
$K^{\ast\pm} \pi^\mp$) result in 4 equations and 4 unknowns
\cite{atwoodtwo}. The four unknowns are the 2 strong phases (one for
each final state), the CP-odd weak phase $\gamma$ that we are after and
the branching ratio (denoted by b) 
of $B^-\to K^-\bar D^0$ which is extremely difficult
to measure experimentally due to severe backgrounds.

One of the advantages of the method is that it can be applied to many
modes e.g.\ $B^-\to K^-,K^{-\ast} D^0,(\bar D^0)$ with $D^0,\bar D^0 \to
K^+\pi^-$, $K^+\rho^-$, $K^{\ast +}\pi^-$, $K^+ a^-_1$,
$K^+\pi^-\pi^+\pi^-$ etc. Another important point is that the method
allows to include $D^0,\bar D^0$ decays to CP-eigenstates
\cite{gronone} so long as one CP-non-eigenstate is also included. So as
a specific example one could use $D^0,\bar D^0\to K^0_s\pi^0$ (i.e.\ a
CPES) with $D^0,\bar D^0\to K^+\pi^-$ (a CPNES).  The point is that
once one CPNES is included sufficient number of observables become
available to solve for the branching ratio $B^-\to K^-\bar D^0$ as an
output along with $\gamma$ \cite{atwoodtwo}.

Fig.~1 illustrates use of only two modes, one CPNES ($K^+\pi^-$) and one
CPES ($K^0_s\pi^0$) of $D^0,\bar D^0$ assuming $\hat N_B=($\# of $B$-$\bar
B$ pairs)${}\times{}$(acceptance)${}=10^8$. Solutions to the equations
for the two modes intersect in four places in the $b$-$\gamma$ plane.
Regions with 68\%, 90\% and 99\% CL are shown. Multiple solutions are
clearly a limitation. Fig.~2 shows the result when several more modes
of $D^0,\bar D^0$  are also combined. Now the improvement over Fig.~1
is significant and $\gamma$ with a (1-sigma) accuracy of about $7^\circ$ is
obtained; in this calculation true value of $\gamma$ is assumed 
to be 60 degrees.


\begin{figure}
\epsfxsize 6 in
\mbox{\epsfbox{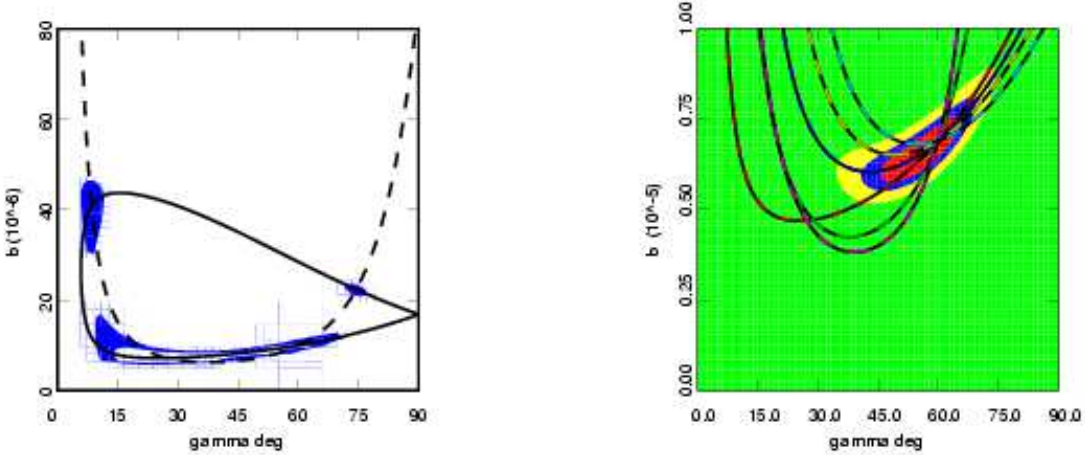}}
\caption{ ~The likelihood distribution is shown
as a function of $\gamma$ and
$b(K^*)$
(which is branching ratio of $B^- \to K^{*-}\bar D^0$) 
assuming that $\hat N_B=10^8$ 
and assuming only the $K^+\pi^-$
and $K_s\pi^0$ modes of $D^0$ are measured.
The outer edge of 
the shaded regions correspond to
$90\%$  confidence
while the inner edge corresponds to
$68\%$  confidence. The solid lines
show
the locus of points which give the $K^+\pi^-$ results while the
short dashed curve shows the points which give the
$K_s\pi^0$ results.
}
\caption{
~The likelihood distribution as in Fig.1 but now using 6 decay modes of $D^0$,$\bar D^0$.  
The solution for the $K^+\pi^-$ data is shown with the solid curve;
that for the $K_S\pi^0$  is shown with 
the short dashed curve;
$K^+\rho^-$ is shown with the long dashed curve;
$K^+a_1^-$ is shown with the dash-dot curve;
$K_s \rho^0$ data is shown with the dash-dot-dot
curve  and
the solution for the $K^{*+} \pi^-$ data is shown with the dash-dash-dot
curve.
}
\end{figure}


\subsection{Extracting $\beta$-$\alpha$ with zero theory error}

The angle $\delta\equiv \beta-\alpha+\pi = 2\beta+\gamma$, can be obtained
by time dependent $\notcp$ asymmetry measurements in $B^0,\bar B^0\to
K^0D^0(\bar D^0)$. As in the case of $B^\pm$ decays to $K^\pm D^0(\bar
D^0)$ only tree graph decays are involved and no penguin, strong or
electroweak, enters. Also once again it is simply a matter of writing down
a bunch of equations and providing enough experimental information via
observables to render the system completely soluble yielding the unitarity
angles that we are after. Indeed there is so much redundancy that not only
$\alpha$ but also $\beta$ can be determined in this way providing a
valuable comparison against the $\beta$ determined from $B\to \psi K^0_s$.

Some aspects of this method have been previously studied by quite a few
authors \cite{bran,kays,sand}. In principle, time-dependent CP asymmetry
(TDCPA) measurements in $B^0(\bar B^0)\to K_sD^0, K_s\bar D^0$ is all that
is needed to extract $\delta$. However, as in the case of $B^\pm\to K^\pm
D^0,(\bar D^0)$, the $D^0,\bar D^0$ flavor tagging is problematic. The
resolution of this problem using CP-non-eigenstates of $D^0,\bar D^0$
(just as in the case of $B^\pm$ decays) has been suggested to give
information on $\delta$ and also possibly on $\beta$ \cite{kays}.

Actually in this case $D^0(\bar D^0)$ decays to CP eigenstates can also be
used. However, if only an exclusive CPES (e.g.\ $K_s\pi^0$) is used then
the number of observables is 3 and the number of unknowns, including
$\delta$, is 4 so not information is available for a separate solution.

Combining the CPES and CPNES methods seems very effective as the number of
unknowns involved for the CPES case are just a subset of those for the
CPNES case. If one exclusive CPNES mode (such as $K^+\pi^-$) and a CPES
mode (e.g.\ $K_s\pi^0$) are used then we get 9 observables for 5 unknowns;
if $\beta$ is treated as an unknown even then the system of equations is
solvable.

Of course several CPNES modes can be included. For each CPNES that is
included we have 6 new observables at the expense of only one additional
unknown. Many exclusive modes are available e.g.\ $K^-\pi^+(Br\sim
3.8\%)$, $K^-\rho^+$ (10.8\%), $K^{\ast-}\pi^+$ (5.0\%), $K^{\ast0}\pi^0$
(3.1\%), $K^{\ast-}\rho^+$ (6.1\%), and $K^{\ast-}a^+_1$ (7.1\%), for a
total of 36\%. A nice 4-body mode with all charged track is
$K^-\pi^+\pi^+\pi^-$ (BR${}>.6\%$).

In fact a very nice way to solve for $\delta$ (and $\beta$ simultaneously)
is to generalize the above exclusive (CPNES) case to inclusive CPNES via
$D^0\to K^-+X$. Then the Br$\sim53\%$; one has 6 observables and 6
unknowns.  So it is a solvable system but with discrete ambiguities. A
very promising way to overcome the ambiguities is to combine this
inclusive CPNES case with a CPES mode.

Fig.~3 illustrates the results of our study of extracting $\delta$ via
this method.  Combining the inclusive CPNES ($D^0\to K^-+X$) with
exclusive CPES seems to do a very good job of eliminating the ambiguities
and give $\delta$ with an error of $\pm2.5^\circ$ (the true value of
$\delta$ in this case study is $110^\circ$). (See also
Table~\ref{tabthree}~\cite{kl}.) 
In this example we have used $\hat N_B={}$(number
of $B$-$\bar B$ mesons) (acceptance)${}=10^9$.

\begin{figure}
\epsfxsize 6 in
\mbox{\epsfbox{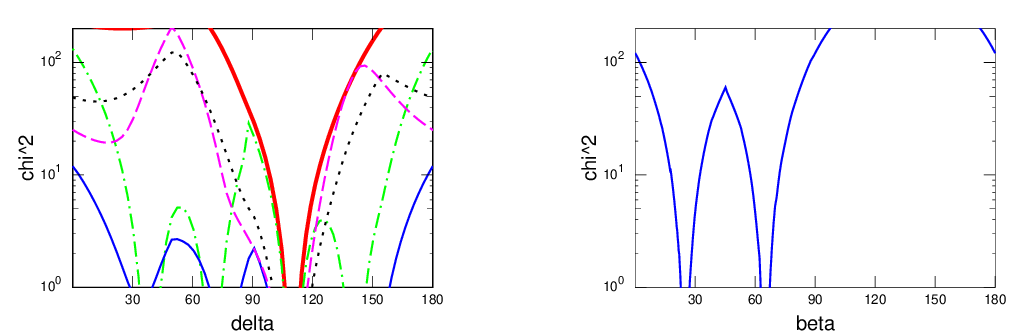}}
\caption{
The $\chi^2_{min}$ vs. $\delta$ for the toy model calculation given
$\hat
N_B=10^9$.  The thin solid line is the result for $D^0\to K^-\pi^+$
alone.  The dashed line the result for CPES containing $K_S$ together
with related CPES containing $K_L$.  The dotted lines the result
obtained
combining $K^-\pi^+$ with CPES containing $K_S$. The dashed-dotted line
gives the result for $K^-+X$ alone and the thick solid line combines
$K^-+X$ with CPES containing $K_S$. (Note
$\delta=110^\circ$ is assumed here)
}
\caption{
The $\chi^2_{min}$ vs. $\beta$ for the toy model calculation given
$\hat
N_B=10^9$
using $K^-+X$ with CPES containing $K_S$. (Note
$\beta=25^\circ$ is assumed.)
}
\end{figure}

A very important feature of this method is that you can also use it to
solve for $\beta$ very cleanly, i.e. no theory assumptions are involved.
Infact this method is even cleaner than the $B \to \psi K_s^0$ method as
the latter does receive some (although very small) penguin contributions
whereas the former has none. However, it is not as efficient as the $\psi
K_s^0$ method, i.e. more number of B mesons are needed to get similar
quantitaive accuracy on $\beta$. Fig.4 illustrates how well the method
works for determining $\beta$. Here the input used is the one given in the
5th row of Table~\ref{tabthree}.  With ${\hat N}_B=10^9$ the one sigma
error on $\beta$ is around 2 degrees.

%
%

\begin{table}[t] 
\vspace*{.5in}
\caption{\emph Attainable one sigma accuracy with 
various data sets
given
${\hat N}_B=10^9$;
note the 2nd and 5th cases are omitted from Fig~3
for clarity. \label{tabthree}
}
\begin{tabular}{|p{2.6 in}|l|}
\hline
Case                                                            &
Accuracy \\
\hline
CPES with $K_S$ and with $K_L$                                  &
$\pm 8.5^\circ$ \\
\hline
CPNES $K^- \pi^+$ with $K_S$ and with $K_L$                     &
$\pm 5^\circ$\\
\hline
The
CPNES $K^- \pi^+$ together with CPES, both with $K_S$ only      &
$\pm 9.0^\circ$ \\
\hline
$K^- + X$ together with $K_S$ CPES                              &
$\pm 2.5^\circ$ \\
\hline
$K^- + X$ together with $K_S$ as well as $K_L$ CPES             &
$\pm 2.4^\circ$\\
\hline
\end{tabular}

\vspace*{.5in}
\end{table}

Table~\ref{tabfour} shows a brief summary for determination of 
the three angles of the
unitarity triangle. It also highlights the limiting theory error of
each method.  With the large data samples that should become available,
it is reasonable to expect that 
with these $B \to K (K^*) D^0$ methods all three angles could be determined
precisely providing an extremely important test of the CKM paradigm.
A notable feature of these methods
is that in charged or neutral B-decays final states relevant to extracting
all three angles cleanly all contain $D^0$ or $\bar D^0$; this should help in
improving the experimental efficiency.

\begin{table}
\begin{center}
\caption{Pristine methods for extracting the UT with
negligible theory error. Lower range of \# of B's is the estimate for initial 
determination and the upper for precise measurments. \label{tabfour}}
\begin{tabular}{l|c|c|c|c|c|c|c}
\hline
& & & & \multicolumn{2}{c|}{} & Limiting & \# of
$B$'s\\
& 
& 
& Type of & \multicolumn{2}{c|}{Pollution} & Theory &  Needed\\
\cline{5-6}
Angle & Mode(s) & Ref. & CP & QCDP & EWP & Error & /$10^8$\\
\hline
 & & & & & & & \\
 & &  Bigi + & & & & & \\
  $\beta$  & $B\to \psi K^0$ &  Sanda  & 
 time dep.  & $\sim1$\% &
  $\lsim1\%$  &  $\sim1$\% &  0.5--5 \\
  $(\phi_1)$  & & & & & & & \\
 & & & & & & & \\
 \hline
 & & & & & & & \\
  $\gamma$  & $\!\!\!\!\!\!K^\pm D^0_\downarrow (\bar
 D^0_\downarrow)$  & Atwood  &  Direct  &  0 
 &  0  &  $\sim0$ &  5--50  \\
  $(\phi_3)$ & \quad \quad$K^+\pi^-$  & 
 Dunietz, & & & & & \\
 & &  Soni & & & & & \\
 & & & & & & & \\
\hline
& & & & & & & \\
$\alpha(\phi_2)$ & $K^0D^0(\bar D^0)$ & Atwood & Direct
& 0 & 0 & $\sim0$ & 5--50 \\
 and & $\downarrow$  & +AS & & & & & \\
$\beta(\phi_1)$ &  CPES,CPNES, & & & & & & \\
&  Inclusive & & & & & & \\
& & & & & & & \\
\hline
\end{tabular}
\end{center}
\end{table}

Recall that two prominent methods for $\alpha$, both using time
dependent CP, have been studied for quite some time. 
The first method~\cite{grontwo},
requires CP asymmetry
measurements in $B^0(\bar B^0)\to\pi^+\pi^-$ as well as BR for
$B^0,\bar B^0\to \pi^0\pi^0$ and $B^\pm\to \pi^\pm\pi^0$.  Although
these measurements should enable one to perform an isospin analysis and
remove the penguin contribution the value of $\alpha$ thus deduced
suffers from contamination from electroweak penguin contribution as EWP
evade the isospin analysis. To that extent this method for $\alpha$
determination has some residual theory error and has to invoke model
dependent estimates in the evaluation of the EWP contribution.

\begin{table}
\caption{Illustrative sample of methods for extracting
the UT. Lower range of \# of B's is the estimate for initial
determination and the upper for precise measurments \label{tabfive}} 
\hspace*{-.5in}\vbox{\begin{tabular}{l|c|c|c|c|c|c|c}
\hline
& & & & \multicolumn{2}{c|}{} & Limiting & \# of
$B$'s\\
& & Ref. & Type of & \multicolumn{2}{c|}{Pollution} & Theory &  Needed\\
\cline{5-6}
Angle & Mode(s) & \  & CP & QCDP & EWP & Error & /$10^8$\\
\hline
 & & & & & & & \\
 & &  Bigi + & & & & & \\
 $\beta$  & $B\to \psi K^0$ &  Sanda  & 
 time dep.  & $\sim1$\% &
  $\lsim1\%$  &  $\sim1$\%&  0.5--5 \\
 $(\phi_1)$ & & & & & & & \\
 & & & & & & & \\
 \hline
 & & & & & & & \\
  $\gamma$  &  $\!\!\!\!\!\!K^\pm D^0_\downarrow (\bar
 D^0_\downarrow)$  &  Atwood  & Direct  &  0 
 &  0  &  $\sim0$  &  5--50  \\
  $(\phi_3)$  & \quad \quad$K^+\pi^-$ & 
 Dunietz, & & & & & \\
 & &  Soni  & & & & & \\
 & & & & & & & \\
\hline
& & & & & & & \\
$\alpha(\phi_2)$ & $K^0D^0(\bar D^0)$ & Atwood &time dep. 
& 0 & 0 & $\sim0$ & 5--50 \\
 and & $\downarrow$  & +AS & & & & & \\
$\beta(\phi_1)$ &  CPES,CPNES, & & & & & & \\
&  Inclusive & & & & & & \\
& & & & & & & \\
\hline
& & & & & & & \\
&  $\pi\pi$  &  Gronau + &  time dep. &  $\approx
30\%$ &  few\%  & $\sim5$--10\%  &  10--50  \\
& &  London  & & & &  & \\
& & & & & & & \\
 $\alpha$  & $\rho\pi$  &  Quinn {\it et al}  & 
{\tt"}  &   $\approx 30\%$  &  {\tt "}  &  $\sim
5$--10\%  &  5--50  \\
 $(\phi_2)$  & & & & & & & \\
&  $\rho(\omega)P$ &  Atwood +   &  {\tt "}  &
 $\approx 30\%$  & {\tt "}  &  $\sim1$--2\%  & 
5--50  \\
& ($P=\pi,$ &  Soni & & & & & \\
&  $\eta,a_0\dots$) & & & & & & \\
& & & & & & & \\
& $\rho\pi+\rho(\omega)P$  &  Comb. of  & {\tt"}
 &  {\tt"}  &  {\tt "} &  $\sim1\%$  & 
5--50  \\
& &  above 2 & & & & & \\
\hline
& & & & & & & \\
 $\alpha(\phi_2)$  &  $B^\pm,B^0_\downarrow(\bar
B^0_\downarrow)$
  &
 Atwood   &  Direct  & 
$\approx 20\%$  &  $\approx 5\%$  &  $\lsim5\%$  &
 5--50  
\\
and&  \qquad \qquad $\rho\omega$, $K^{\ast0}\rho^+$  & +AS  & & &
& & 
\\
& & & & & & & \\
 $\gamma(\phi_3)$ & $B\to
K^\ast\rho(\omega)$ &  {\tt "}  &  {\tt "}  & 
{\tt "}  &  {\tt "}  &  {\tt "}  &  5--50  \\
\hline
\end{tabular}}
\end{table}

The key experimental difficulty in this set of measurements is the
$2\pi^0$ mode due to the small branching ratio ($\sim2\times10^{-7}$)
that is expected, made harder by the relative low detection efficiency,
perhaps also compounded by the fact that the $\pi^0$'s are very
energetic giving rise to a small opening angle between the
$\gamma$-pairs. Presumably, these experimental difficulties will be
surmounted as luminosities improve. Already  the two groups have made
attempts to measure TDCPA in $B^0\to\pi^+\pi^-$ which should become
quite accurate relatively shortly. However, the interpretation of this
in terms of the angle $\alpha$ requires important input from theory. 

Another important method wherein isospin analysis, 
can be used is $B\to \rho\pi$ \cite{snyd}. In this approach one can
make use of resonance effects in the Dalitz plot; however, some model
dependence is likely to occur in handling the continuum of $B\to 3\pi$.
Once again, EWP are assumed to be negligible and this is also an
important source of the limiting theory error in the $\rho\pi$
analysis.

Recently we have also proposed two other methods
\cite{atwoodfour,atwoodfive} for extracting $\alpha$ and $\gamma$ that use
penguin and tree interference effects and therefore are also not
completely clean. Table \ref{tabfive} shows a sample of these methods that
uses penguin-tree interference along with the very clean methods of Table
\ref{tabfour}. It is also important to emphasize that $B \to K D^0$
methods require roughly the same number of B's as the $\rho \pi$ or $\pi
\pi$ methods.

While our emphasis here has been on $B^+,B^-,B^0,\bar B^0$ mesons, if time
dependent oscillation studies in $B_s,\bar B_s$ become feasible, then a
clean way to get $\gamma$ may also be accessible via final states of
$D_s^+(D_s^-)K^-(K^+)$ \cite{adk}, or their vector
counterparts~\cite{lss}. Experimental feasibilty especially
in a hadronic environment of some of these is studied 
in \cite{ss}.

Let us briefly remark that there is also a very 
clean way to get the magnitude of the CP-odd CKM phase directly 
from $K_L \to \pi^0 \nu \bar\nu$~\cite{ab2,sk2}.

\section*{ Acknowledgements} 
{
We thank Jean-Marie Frere and Tran
Thanh Van for the invitation. 
This
research was supported by the US DOE  contract
\#DE-AC02-98CH10886(BNL) and DE-FG02-94ER40817(ISU).}


\begin{thebibliography}{99}
\bibitem{zero} The phrase zero theory error will be used here loosely
to emphasize negligible theory error i.e. $\lsim 0.1\%$.
\bibitem{notation} Note that there are two notations in use,
$(\alpha,\beta,\gamma) \equiv (\phi_2,\phi_1,\phi_3)$
\bibitem{kl} With regard to $K_L$ in this Table, note that their
detection is not necessary for this method to work. However, 
they can aid the analysis considerably as they add to the number
of observables; see \cite{atwoodthree}. For numerical
results in this Table we assumed $K_L$ detection efficiency
to be half of that for $K_S$.
\end{thebibliography}
\end{document}